\documentclass[review,12pt]{elsarticle}%
\usepackage{amsfonts}
\usepackage{color}
\usepackage{amsmath}
\usepackage{amssymb}
\usepackage{graphicx}%
\providecommand{\U}[1]{\protect\rule{.1in}{.1in}}
\journal{Physica A}
\begin{document}
%
%TCIMACRO{\TeXButton{Begin frontmatter}{\begin{frontmatter}}}%
%BeginExpansion
\begin{frontmatter}%
%EndExpansion
%

%TCIMACRO{\TeXButton{Title}{\title
%{A 3-states magnetic model of binary decisions in sociophysics}}}%
%BeginExpansion
\title{A 3-states magnetic model of binary decisions in sociophysics}%
%EndExpansion
%

%TCIMACRO{\TeXButton{Author}{\author[mymainaddress]{Miguel A. Fernandez}
%}}%
%BeginExpansion
\author[mymainaddress]{Miguel A. Fernandez}
%EndExpansion
%

%TCIMACRO{\TeXButton{Author}{\author
%[mymainaddress,mysecondaryaddress]{Elka Korutcheva}}}%
%BeginExpansion
\author[mymainaddress,mysecondaryaddress]{Elka Korutcheva}%
%EndExpansion
%

%TCIMACRO{\TeXButton{Author}{\author
%[mymainaddress]{F. Javier de la Rubia\corref{mycorrespondingauthor}}
%\cortext[mycorrespondingauthor]{Corresponding author}
%\ead{jrubia@fisfun.uned.es}}}%
%BeginExpansion
\author[mymainaddress]{F. Javier de la Rubia\corref{mycorrespondingauthor}}
\cortext[mycorrespondingauthor]{Corresponding author}
\ead{jrubia@fisfun.uned.es}%
%EndExpansion
%

%TCIMACRO{\TeXButton{Address}{\address[mymainaddress]{Departamento de F\'{\i
%}sica Fundamental, Universidad Nacional de Educaci\'{o}%
%n a Distancia (UNED), Paseo Senda del Rey 9, E-28040 Madrid, Spain}
%}}%
%BeginExpansion
\address[mymainaddress]{Departamento de F\'{\i}%
sica Fundamental, Universidad Nacional de Educaci\'{o}%
n a Distancia (UNED), Paseo Senda del Rey 9, E-28040 Madrid, Spain}
%EndExpansion
%

%TCIMACRO{\TeXButton{Address}{\address
%[mysecondaryaddress]{Also at G. Nadjakov Inst. of Solid State Physics, Bulgarian Academy
%of Sciences, 1784 Sofia, Bulgaria.}}}%
%BeginExpansion
\address
[mysecondaryaddress]{Also at G. Nadjakov Inst. of Solid State Physics, Bulgarian Academy
of Sciences, 1784 Sofia, Bulgaria.}%
%EndExpansion
%

%TCIMACRO{\TeXButton{Begin abstract}{\begin{abstract}}}%
%BeginExpansion
\begin{abstract}%
%EndExpansion

We study a diluted Blume-Capel model of 3-states sites as an attempt to
understand how some social processes as cooperation or organization happen.
For this aim we study the effect of the complex network topology on the
equilibrium properties of the model, by focusing on three different
substrates: random graph, Watts-Strogatz and Newman substrates. Our computer
simulations are in good agreement with the corresponding analytical results.%

%TCIMACRO{\TeXButton{End abstract}{\end{abstract}}}%
%BeginExpansion
\end{abstract}%
%EndExpansion
%

%TCIMACRO{\TeXButton{Begin keyword(s)}{\begin{keyword}}}%
%BeginExpansion
%EndExpansion
%

%TCIMACRO{\TeXButton{End frontmatter}{\end{frontmatter}}}%
%BeginExpansion
\end{frontmatter}%
%EndExpansion

\section{Introduction}

There is great interest in applying physical models of proven efficiency in
the new field of sociophysics. Problems such as decision of living in a
neighborhood \cite{Schelling}, to go to a crowed bar \cite{Arthur} or to
participate in a strike \cite{Gefen} are some particular examples of such applications.

Physics ideas in social science have been introduced in order to study
hierarchical structures by using the principle of least difficulty
\cite{Toulouse}, the effect of frustration for modelling the dissemination of
culture within the Axelrod model in his energy landscape theory
~\cite{Axelrod}, and by Galam in his models of coalitions~\cite{Galam}. To
study the emergence of order in such a system, magnetic models of the Ising
family have been used~\cite{Doro0,Doro2,Baxter}. In particular, within the
social sciences, models have been developed to describe urban segregation,
language change~\cite{Nettle}, business confidence and economic
opinions~\cite{Stauffer}. These studies apply either the traditional
two-states Ising model to binary choice or a Potts multistates
model~\cite{Galam2} to multiple choice. Representing opinions or positioning
by discrete values, society can be modeled as a system in which an analogue of
the magnetic spin variable represents the state of individuals while the
couplings represent their interactions~\cite{Yang,Parongama}.

The Ising model is usually applied in the context of binary decisions, which
can be to vote or not to vote, to buy or not to buy a certain good
\cite{Nadal1, Nadal2}, etc. The Random Field Ising Model (RFIM) was first
applied to social systems for studing collective phenomena such as consensus
and attitude changes in groups \cite{Galam3} and rational group decision
making \cite{Galam4}. Another useful application of the RFIM in social and
economics sciences is presented in \cite{Bouchaud}, by taking into
consideration heterogeneities and interaction in decision making, thus
providing a unified framework to account for many collective socioeconomic
phenomena leading to ruptures or crisis. Although in the mean-field
approximation, quantitative details of the model might be sensitive to the
type of the topology or the distribution of the idiosyncratic fields, the
qualitative behavior does not depend very much on them \cite{Bouchaud}.

By using coupled Ising models, the interdependent binary choices under social
influence for homogeneous unbiased populations have been analyzed \cite{Ana}.
Currently a comparison of the model with real data of the European labor
market is under investigation \cite{Ana1}.

Important results have also been obtained in \cite{Michard-Bouchaud}, where
real data concerning the drop of birth rates in European countries in the
second half of the XXth century, the increase of cell phones in Europe in the
90s and the way clapping dies out at the end of a concert have been
successfully explained by using the RFIM.

Concerning three-states models, several real examples can be given, such as
the decision to vote, not to vote or to abstain, to belong to NATO, to the
Warsaw Alliance or to a third part Alliance, and so on. For example,
Ref.~\cite{Gekle} studies the opinion dynamics in a three-choice system
dividing the population in random groups of fixed size, demonstrating that
such a system always reaches an equilibrium, and in~\cite{Crokidakis} there is
a discussion on the opinion formation in a voter-like model of a 3-states
agents system (yes, no, undecided), defined on a regular lattice.

The review ~\cite{Castellano} provides many examples of applications of
statistical physics methodology and concepts for the study of social systems,
including opinion formation, cooperation, cultural dynamics, language
evolution, crowd behavior or human dynamics among many others.

On the other hand, modeling of Small World and scale-free networks allowed for
powerful theoretical analyses of dynamical collective social phenomena, such
as opinion formation, infection, or damage propagation in social networks,
where the analysis of social structures and interactions is crucial. Both
models can be combined to study dynamical processes on complex
networks~\cite{Doro2,Doro,Fernandez}.

Activities like cooperation and coalition forming have been traditionally
approached by game theory in politics and diplomacy~\cite{Antal}, yet it is
only recently that the concepts of Statistical Physics have been applied. In
this context, various investigations have been done by representations in
terms of cellular automata,~\cite{Sefer,Sefer2} or in terms of magnetic models
on a Cayley tree, similar to what we propose in this work~\cite{Gani,Gani2}.
In this article we demonstrate that starting from a rather weak assumption,
stating that each individual in the ensemble is an agent who can adopt three
different states, we show that it is possible to effectively model a broad
range of sociological processes. We model the society as a set of agents whose
interactions attempt to minimize their frustration. In order to introduce the
neutrality as a possible decision option, different to the symmetric choice,
we use the Blume-Capel model~\cite{BEG,Plischke}. Further, we make our model
reside on an underlying complex network, which plays the role of substrate,
and study the effects of its topology on the model behavior.

The Blume-Capel model has been applied in the context of urban segregation for
the case of a fixed number of agents on a lattice \cite{Schelling1} and for
the case of an external reservoir of agents, thus modelling an open city
\cite{Schelling2}. Although the interpretation of the variables is different,
in the sense that in the problem of opinion formation studied here the zero
state corresponds to neutrality, while in the above cited references they
correspond to an empty location, there are similarities among these models
that will be commented later.

In this paper we will restrict ourselves to the mean-field case, in the sense
that the agents perceive the opinion of the rest of the agents through the
average opinion, thus the total \textquotedblleft demand\textquotedblright%
\ becomes a public information that influences the individual agent.

The rest of the paper is organized as follows: in Section 2 we study magnetic
models in which three states are allowed, in contrast with traditional Ising
models; in Section 3 we study the network on which our model will be residing,
with emphasis on the Annealed Network Approximation approach, which enables us
to derive expressions for the order parameters on the grid substrate; Section
4 is devoted to the study of the order parameters and critical temperature of
our three-states magnetic model on three different network substrates, with a
special focus on the Newman substrate, which is deemed the most realistic.
Finally, in Section 5 we discuss the results.

\section{Three-states magnetic models}

\label{sec:BEG} Blume, Emery and Griffiths (BEG) proposed the first version of
their three-states model in~\cite{BEG} to explain the behavior of a ${}%
^{3}He-{}^{4}He$ mixture, which has been since successfully applied to a
number of different problems.

The three states BEG model has been applied in the context of neural
networks~\cite{Yedidia,Busquets,Korut} giving way to enhanced techniques for
information storage. It has been studied by a number of techniques:
Bethe-Peierls approximation~\cite{Du,Goveas}, real-space
normalization~\cite{Bak} and exact recursion~\cite{Alba1,Alba2,Ekiz,Keskin},
among many others. In this paper we will use the mean-field approximation.

In analogy with the original model, we will consider an ensemble made up by
agents of two species, neutral with spin 0 and non neutral with spin $\pm1$,
defining thus the following two parameters:
\begin{equation}
\mathcal{M}=\frac{1}{N}\sum\limits_{i=1}^{N}{\left\langle {{S}_{i}}
\right\rangle }%
\end{equation}
and
\begin{equation}
x=1-\left\langle S_{i}^{2} \right\rangle ,
\end{equation}
where $\mathcal{M}$ and $x$ represent the number of "active" agents ($S_{i} =
\pm1$) and the neutral agents, respectively, being $S_{i}$ the site spin
(agent state). The Hamiltonian of this system has the general form
\begin{equation}
\label{eq:ham}H =-J\sum\limits_{i,j}{{{S}_{i}}{{S}_{j}}}-\kappa\sum
\limits_{i,j}{S_{i}^{2}S_{j}^{2}+\Delta\sum\limits_{i}{S_{i}^{2}}}.
\end{equation}

Here $J$ is the coupling constant, $\kappa$ is the interaction between agents
of different species and $\Delta$ is the anisotropy term. The limit
$\Delta\rightarrow-\infty$ corresponds to the Ising case, or an absence of
neutral agents, while large values of $\Delta$ corresponds to large density of
neutral agents. If the interactions between agents are equivalent, the
biquadratic term can be neglected, thus obtaining a simplified version of the
Blume-Emery-Griffiths model known as Blume-Capel model, described with detail
in~\cite{Plischke}. We will follow this simplification and adopt the
Blume-Capel model to capture the essence of competition between two processes,
one favoring confrontation and another favoring neutrality. The dilution
manifests itself as random coupling strengths between nodes, and has no effect
in the anisotropy term, which corresponds to self-interaction. Consequently,
we generate in our simulation $J$ from a Poisson distribution, obtaining
results in good agreement with the theory.

\begin{figure}[ptb]
\centering
\includegraphics[scale=0.4]{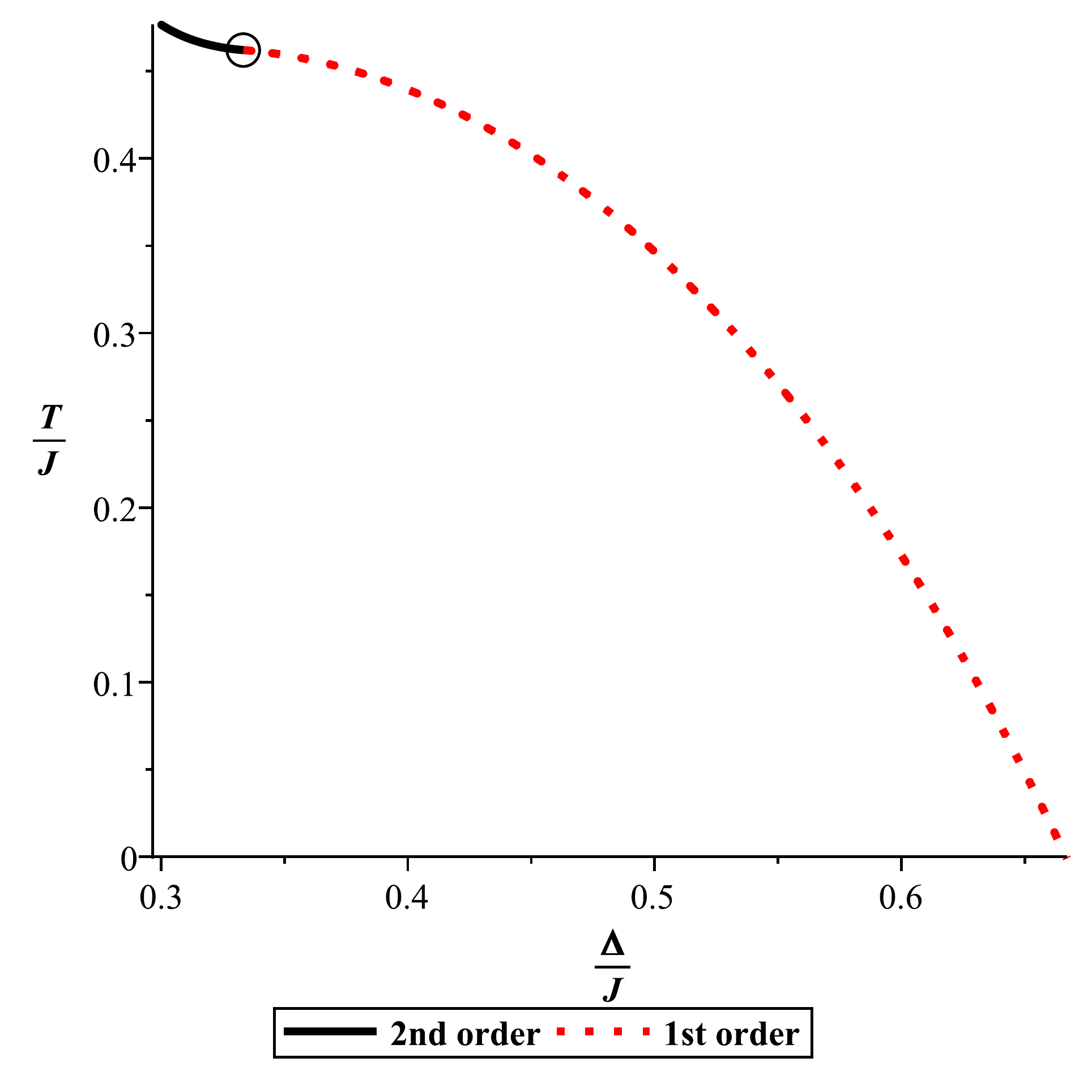}\caption{(Color online).
Tricritical point in Blume-Capel model.}%
\label{BCTCP}%
\end{figure}

We start from the Hamiltonian in~(\ref{eq:ham}) with $\kappa=0$ and notice it
is convenient to work with $1-x$ instead $x$. We then apply the Mean Field
(MF) approach~\cite{Plischke} by splitting the site spin in two parts, the
thermal average and the fluctuation, i.e., as ${{S}_{i}}=\left\langle {{S}%
_{i}} \right\rangle +\delta{{S}_{i}}=m+\delta{{S}_{i}}$, obtaining thus the
following MF Hamiltonian%

\begin{equation}
{{\hat{H}}_{MF}}=\frac{1}{2}NJ_{0}{{m}^{2}}-J_{0}m\sum\limits_{i}{{{S}_{i}}%
}+\Delta\sum\limits_{i}{S_{i}^{2}},
\end{equation}
where we introduced $J_{0}$, the discrete Fourier transform of $J(\vec{R})$
for wavevector $q=0$. After rescaling by $J_{0}$ as
\begin{equation}
\theta=\frac{{{k}_{B}}T}{J_{0}},\delta=\frac{\Delta}{J_{0}},
\end{equation}
we obtain the following expressions for the partition function and the
normalized free energy respectively
\begin{equation}
Z={{e}^{-\frac{1}{2}\beta NJ_{0}{{m}^{2}}}}{{\left[  1+2{{e}^{-\frac{\delta
}{\theta}}}\cosh\left(  \frac{m}{\theta}\right)  \right]  }^{N}},
\end{equation}%
\begin{equation}
f=\frac{F}{NJ_{0}}=\frac{1}{2}{{m}^{2}}-\theta\ln\left[  1+2{{e}%
^{-\frac{\delta}{\theta}}}\cosh\left(  \frac{m}{\theta}\right)  \right]  .
\end{equation}
Finally, minimizing with respect to $m$ and $x$ respectively, we get the order
parameters
\begin{equation}
m=\frac{2{{e}^{-\frac{\delta}{\theta}}}\sinh\left(  \frac{m}{\theta}\right)
}{1+2{{e}^{-\frac{\delta}{\theta}}}\cosh\left(  \frac{m}{\theta}\right)  }
\label{BCM}%
\end{equation}
and
\begin{equation}
1-x=\frac{2{{e}^{-\frac{\delta}{\theta}}}\cosh\left(  \frac{m}{\theta}\right)
}{1+2{{e}^{-\frac{\delta}{\theta}}}\cosh\left(  \frac{m}{\theta}\right)  }.
\end{equation}
This system exhibits a critical temperature at
\begin{equation}
{{\theta}_{c}}=\frac{2}{\exp(\frac{\delta}{{{\theta}_{c}}})+2}%
\end{equation}
and a tricritical point separating first order from second order transition
located at
\begin{equation}
\theta=\frac{1}{3},\delta=\frac{2}{3}\ln2,
\end{equation}
shown in Figure~\ref{BCTCP}.

\section{Annealed Network Approximation}

\label{GS}

In this section and afterwards we will use the annealed MF approach by
accounting for the heterogeneity of a complex
network~\cite{Doro0,Doro2,Doro,Giauranic,Bianconi}.

The main idea is to replace a model on a complex network by a model on a
weighted fully connected graph. This approximation is exact in the limit of
large systems for a Small World network~\cite{Newman}. Let us consider our
magnetic model described by the Hamiltonian obtained in Section~\ref{sec:BEG}
which resides on a graph with a given topology, that will be defined in the
next sections. The corresponding adjacency matrix ${{a}_{ij}}$ will be
replaced by an effective one, thus introducing the probability that nodes $i$
and $j$ with degrees ${{q}_{i}}$ and ${{q}_{j}}$ are connected with weights
$\frac{{{q}_{i}}{{q}_{j}}}{\left\langle q\right\rangle N}$, being
$\left\langle q\right\rangle $ the average degree.

In other words, we substitute the original network with an effective substrate
where the sum of couplings for each node $i$ has the same value $J{{q}_{i}}$
as in the original graph. In this way the fully connected graph approximates
the original complex network.

Using the Hamiltonian~(\ref{eq:ham}) with $\kappa=0$
\begin{equation}
H=-\sum\limits_{i,j}{{{J}_{ij}}{{S}_{i}}{{S}_{j}}}+\Delta\sum\limits_{i}%
{S_{i}^{2}}. \label{ANA1}%
\end{equation}
we can write the corresponding node energy, in the Mean Field Approximation,
as follows:
\begin{align}
&  {{E}_{i}}=H({{S}_{i}})=-{{S}_{i}}\left(  \sum\limits_{j}{{{J}_{ij}}{{S}%
_{j}}}\right)  +\Delta S_{i}^{2}\nonumber\\
&  \approx-{{S}_{i}}\sum\limits_{j}{{{J}_{ij}}{{m}_{j}}}+\Delta S_{i}^{2},
\end{align}
where ${{J}_{ij}}$ are the coupling constants between node $i$ and its
neighbors $j$, ${{m}_{j}}$ is the magnetization of node $j$ and $\Delta$ is
the anisotropy term.

The Annealed Network Approximation described above gives the following
approximations for the order parameters (see the Appendix for the details)%

\begin{equation}
\label{mi}{{m}_{i}}=\frac{2\sinh\beta J{{q}_{i}}M}{{{e}^{\beta\Delta}}%
+2\cosh\left[  \beta J{{q}_{i}}M \right]  }%
\end{equation}

and%

\begin{align}
&  1-x=\frac{2\cosh\beta J{{q}_{i}}M}{{{e}^{\beta\Delta}}+2\cosh\left[  \beta
J{{q}_{i}}M \right]  }.
\end{align}

Here $J$ is the expected value of the coupling constants and we have
introduced the weighted magnetic moment, defined as
\begin{equation}
\label{defM}M=\sum\limits_{j}{\frac{{{q}_{j}}{{m}_{j}}}{\left\langle q
\right\rangle N}}.
\end{equation}
$M$ is a solution of the following equation:
\begin{equation}
\label{M}M=\sum\limits_{q}{\frac{qP(q)}{\left\langle q \right\rangle }}%
\frac{2\sinh\beta JqM}{\exp\left(  \beta\Delta\right)  +2\cosh\beta JqM}%
\end{equation}
and is obtained after substituting Eq.~(\ref{mi}) into the definition of $M$
and by using the fact that $\sum\nolimits_{i}{f({{q}_{i}})=\sum\limits_{q}%
{P(q)f(q)}}$, being $P(q)$ the degree distribution of the original
network~\cite{Doro0,Doro2,Doro}. As can be seen from Eq~(\ref{M}), the order
parameter in our model is the weighted magnetic moment $M$.

\section{Magnetic Models on Complex Networks}

Complex networks represent better the topologies found in real world than
regular grids. In particular, \textit{Small Worlds} networks are often used to
describe economic and social organizations~\cite{Barrat}.

To generate a complex network, Watts and Strogatz add disorder to a regular
grid an thus get a \textit{Small World} analytic model which combines high
clustering and short paths~\cite{Watts2,Watts}. The rationale behind is that a
pure random network ignores the clustering, whereby nodes connected to a node
are often also connected between themselves. In this way, a Watts-Strogatz
network reunites local properties of a regular grid with global properties of
a random network, thanks to the introduction of long range connections in an
otherwise initially regular grid. In the original variation of the
Watts-Strogatz network, as described in~\cite{Barrat} and~\cite{Watts}, the
network starts from N nodes located on an one-dimensional grid, with links to
$z$ next neighbors. Then, a link between tho nodes is selected randomly with
probability $\phi$ and replaced by a shortcut from the first to a third random
node. In a second variation, due to Newman~\cite{Newman,Newman2}, shortcuts
are just added without deleting the original links.

We postulate that our model resides in a grid described by a Small-World type
network, for which we will use three different models: a random graph and the
two variations of the Watts-Strogatz model described above, the original by
Watts and Strogatz and the variation proposed by Newman. In this section, we
will apply the annealed network approximation to a Blume-Capel system residing
in these three different diluted network substrates. For that purpose, the
weighted neutrality and magnetization are derived using in each case the
relevant degree distribution.

\subsection{Random Graph Substrate}

The random graph is the first model utilized to study complex networks. In it,
randomly chosen nodes are connected, so that in a certain way, a random graph
model is complementary to a lattice type model like the ones we will study in
following sections. Its degree distribution shows a rapid drop at large
degrees, so that we can safely consider a cutoff of 20 for the sum obtained
substituting the distribution in Eq.~(\ref{M}). Therefore we get the following
expression for the weighted magnetic moment
\begin{equation}
M=\sum\limits_{q=1}^{20}{\frac{q}{\left\langle q \right\rangle }}%
{{e}^{-\left\langle q \right\rangle }}{{\left\langle q \right\rangle }^{q}%
}\frac{1}{q!}\frac{2\sinh\beta JqM}{\exp\left(  \beta\Delta\right)
+2\cosh\beta JqM}.
\end{equation}
We will see in next Section (\ref{WS}) that Watts-Strogatz model is not
defined for $q < 6$, hence we make $\left\langle q \right\rangle =6$, so that
it agrees with the values used to study the other topologies. Then we have the
following order parameters
\begin{equation}
\label{M0}M=\sum\limits_{q=1}^{20}{q}{{e}^{-6}}{{6}^{q-1}}\frac{1}{q!}%
\frac{2\sinh\beta JqM}{\exp\left(  \beta\Delta\right)  +2\cosh\beta JqM}%
\end{equation}
and
\begin{equation}
1-x=\sum\limits_{q=1}^{20}{q}{{e}^{-6}}{{6}^{q-1}}\frac{1}{q!}\frac
{2\cosh\beta JqM}{\exp\left(  \beta\Delta\right)  +2\cosh\beta JqM}.
\end{equation}

\begin{figure}[ptb]
\centering
\includegraphics[width=\textwidth]{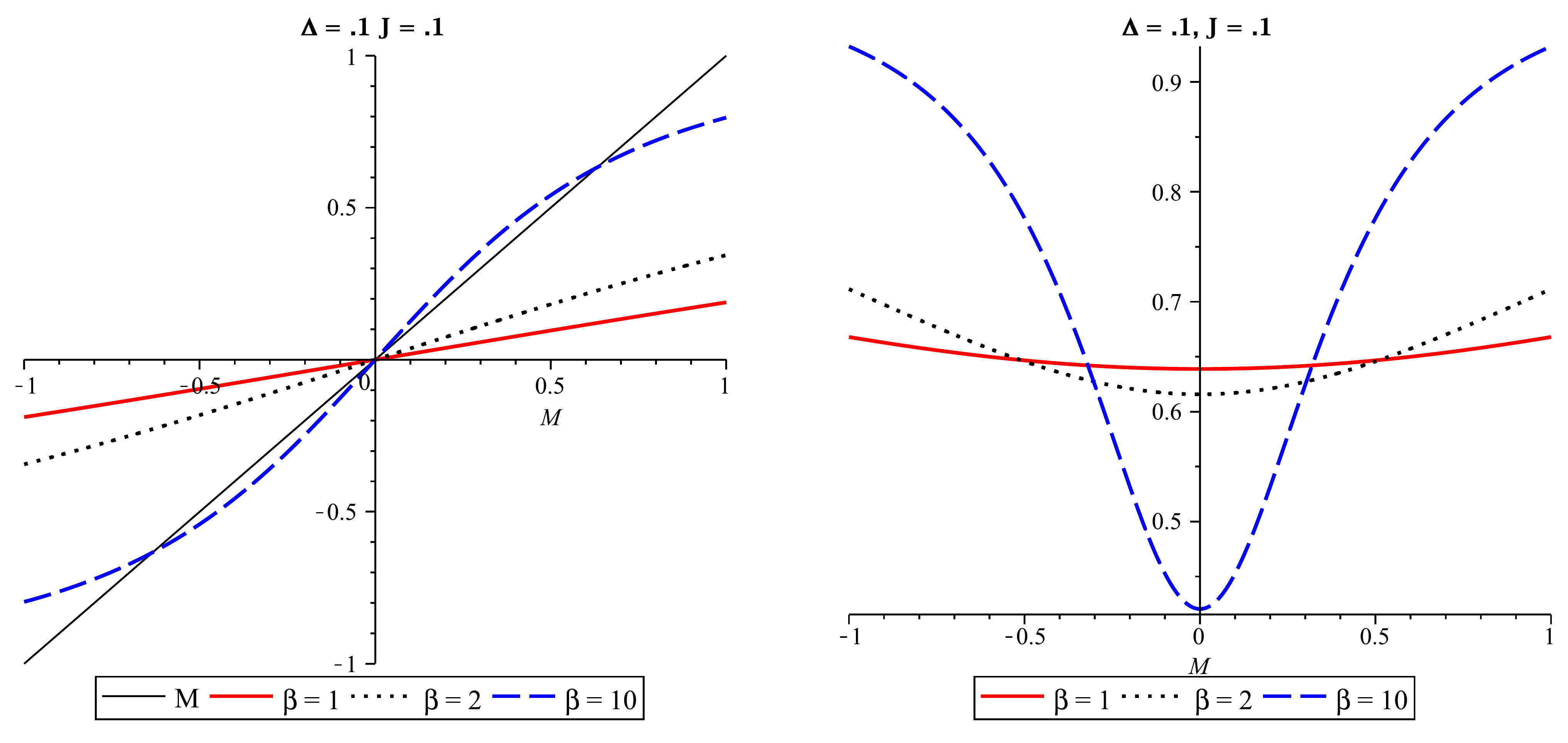}\caption{(Color
online). Solutions for magnetization (left) and neutrality (right) in Random
Graph substrate.}%
\label{RGframe}%
\end{figure}

Figure~\ref{RGframe} shows solutions of Eq.~(\ref{M0}), being $J$ the expected
value of a normal distribution for the quenched coupling constants. $M= 0$ is
always a solution, and there is a limiting value of $\Delta$ for which it is
unique. Below this value only exists the disordered state and above there are
more finite solutions. A very small connectivity causes also $M= 0$ to be the
only solution for every value of $\Delta$ and temperature. Either very low
connectivity -because there is no interaction- or high propensity to
neutrality prevent collective decisions.

The critical temperature at which a non-zero value of $M$ is possible is found
graphically looking at the intersections of $y=M$, represented by a solid line
in the left of Figure~\ref{RGframe} and the rhs of Eq.~(\ref{M0}) to get:
\begin{equation}
\frac{1}{{{\beta}_{c}}}=\frac{CJ}{{{e}^{{{\beta}_{c}}\Delta}}+2},
\end{equation}
where $C$ depends on the connectivity. The meaning of the existence of a
critical temperature is that a very high agitation, either thermal as in the
magnetic model or social, prevents the agents to settle in a fixed state. With
respect to the order parameter related with neutrality, when $\Delta=0$ and
$M=0$ all solutions of $1-x$ are null at any temperature, i.e., all the agents
are neutral.

\subsection{Watts-Strogatz Substrate}

\label{WS} To build their model, Watts and Strogatz start from a regular
unidimensional grid in which each node is connected to $2z$ next neighbors.
Then links are selected randomly with probability $\phi$ and one end is
reconnected to another randomly selected node.

To get the distribution of connectivity $P(q)$, see~\cite{Barrat}, we realize
that by network construction, the average connectivity is $K=2z$. The
connectivity of a node can be written as ${{q}_{i}}=z+{{n}_{i}}$, with
${{n}_{i}}\geq0$. Now, ${{n}_{i}}$ can be split in two parts: $n_{i}^{1}\leq
z$ remaining links (with probability $1-\phi$) and $n_{i}^{2}={{n}_{i}}%
-n_{i}^{1}$ links which have been reconnected to node $i$ with probability
$\frac{\phi}{N}$ obtaining
\begin{equation}
P(n_{i}^{1})=\dbinom{z}{n_{i}^{1}}{{(1-\phi)}^{n_{i}^{1}}}{{\phi}^{z-n_{i}%
^{1}}}%
\end{equation}
and
\begin{equation}
P(n_{i}^{2})=\frac{{{(z\phi)}^{n_{i}^{2}}}}{n_{i}^{2}!}\exp(-\phi z),
\end{equation}
for large N. Finally
\begin{equation}
P(q,\phi,z)=\sum\limits_{n=0}^{\min(q-z,z)}\dbinom{z}{n}{{{(1-\phi)}^{n}%
}{{\phi}^{z-n}}\frac{{{\left(  \phi z\right)  }^{q-z-n}}}{\left(
q-z-n\right)  !}}\exp\left(  -\phi z\right)  \label{PWS}%
\end{equation}

As this expression is valid for $z\geq3$ (otherwise we could have finite
probability for degrees $<0$), we will take $z=3$ in the rest of this section.

Summarizing, the degree distribution is expressed in the following way:

\begin{enumerate}
\item For $\phi=0$, the distribution is $\delta(x-K)$. This case corresponds
to the original grid without shortcuts, so that its distribution is a Dirac's
delta centered on the average value, $\left\langle q \right\rangle =2z$ as all
the nodes have the same connectivity.

\item For $\phi=1$ we have a Poisson distribution $P(q)={\exp({-\left\langle
q\right\rangle })}{{\left\langle q\right\rangle }^{q}}\frac{1}{q!}$. This is
an Erd{\"{o}}s-Renyi network, historically the first used to study complex
networks. It is analytically easy to deal with, but not too representative of
real networks.

\item For values of $\phi$ between 0 and 1, which better approximates real
networks we have Eq.~(\ref{PWS}).
\end{enumerate}

\begin{figure}[ptb]
\centering
\includegraphics[scale=0.5]{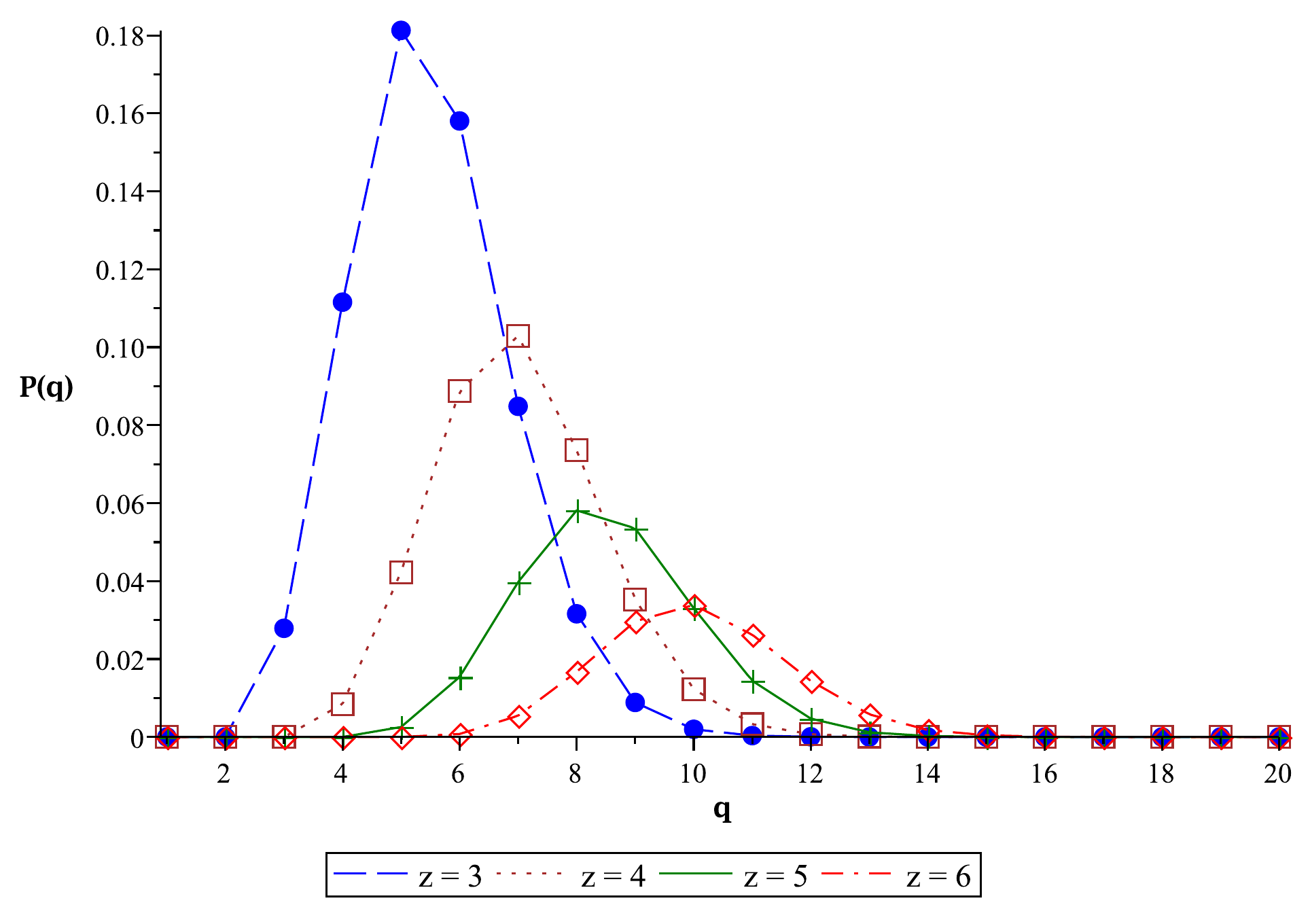}\caption{(Color online).
Connectivity in Watts-Strogatz substrate for different levels of
neighborhood.}%
\label{WSzs}%
\end{figure}

Here, $\phi$ is the reconnection probability, $z$ is the level of separation
of nearest neighbors and $\left\langle q \right\rangle =2z$ is the
connectivity. Figure~\ref{WSzs} shows the connectivity distribution for
various values of neighborhood levels. We see that, unlike real networks, this
distribution is peaked with low cutoff, with width depending only on $\phi$.
But then, this model was conceived to reproduce the clustering and short path
lengths observed in real networks, not their connectivity.

Observing the degree distribution in Figure~\ref{WSzs} we see a rapid drop at
large degrees, as in the random graph, therefore we use a cutoff of 20. Then
the Annealed Network approximation, applied to the medium range of the
reconnection probability, yields for the order parameters%

\begin{equation}
M=\sum\limits_{q=3}^{20}{\left\{  \frac{q}{{\left\langle q\right\rangle }%
}\cdot P(q,\phi,z)\cdot\frac{2\sinh\beta JqM}{\exp\left(  \beta\Delta\right)
+2\cosh\beta JqM}\right\}  }\label{M1}%
\end{equation}
and
\begin{equation}
1-x=\sum\limits_{q=3}^{20}{\left\{  \frac{q}{{\left\langle q\right\rangle }%
}\cdot P(q,\phi,z)\cdot\frac{2\cosh\beta JqM}{\exp\left(  \beta\Delta\right)
+2\cosh\beta JqM}\right\}  }.
\end{equation}

\begin{figure}[ptb]
\centering
\includegraphics[width=\textwidth]{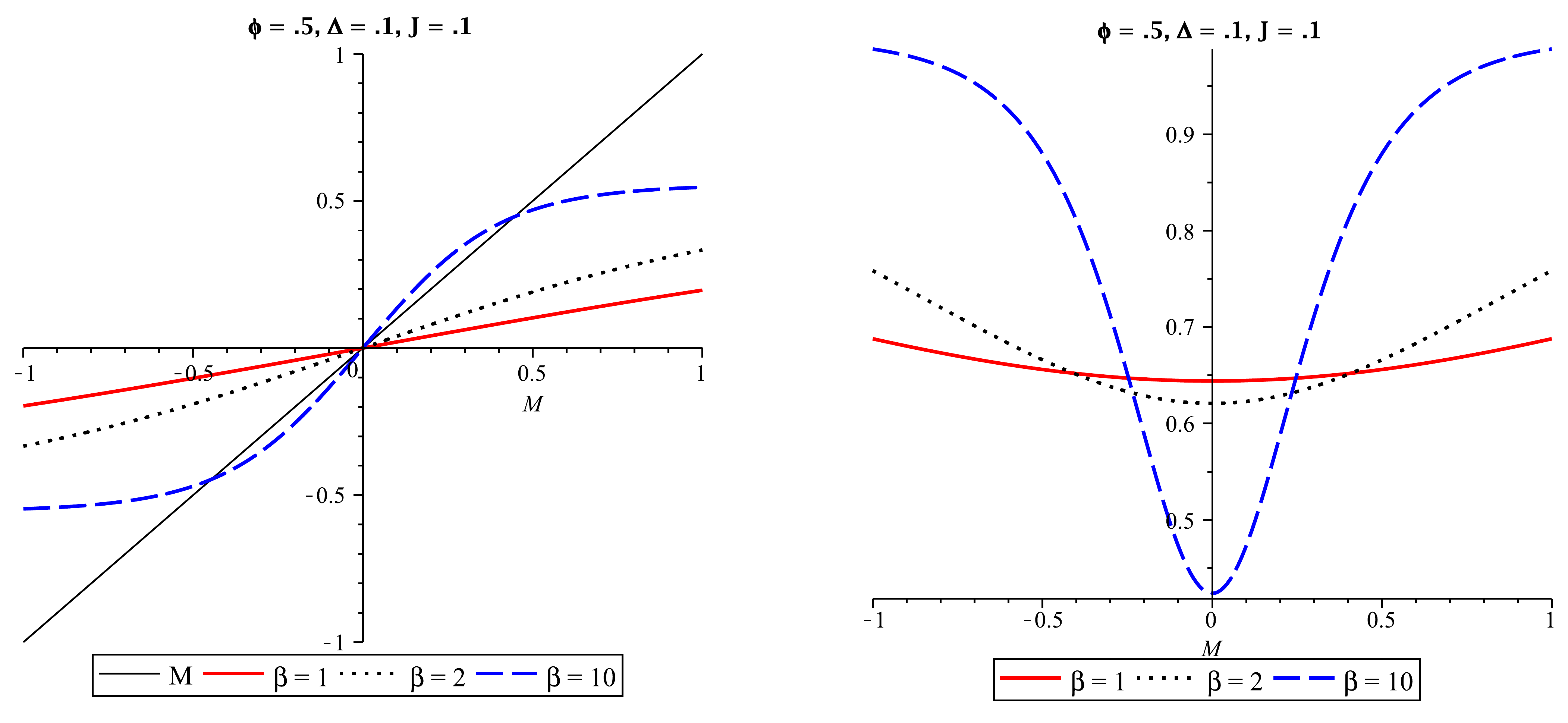}\caption{(Color
online).Solutions for magnetization (left) and neutrality (right) in
Watts-Strogatz substrate.}%
\label{WSframe}%
\end{figure}

As for the case of Poisson, the disordered state characterized by $M=0$ is
always a solution, and below certain value of $\Delta$ it is unique. The
critical temperature is found graphically as before, looking at the
intersections of $y=M$, represented by a solid line in the left of
Figure~\ref{WSframe}, and the rhs of Eq.~(\ref{M1}). In this way we get the
critical value of $\beta$ for which a non-zero value of $M$ is possible:
\begin{equation}
\frac{C{{\beta}_{c}}J}{{{e}^{{{\beta}_{c}}\Delta}}+2}=1.\label{TC}%
\end{equation}

Here $C$ is a constant which depends approximately on the reconnection
probability $\phi$ by a potential law of the form
\begin{equation}
C={{C}_{0}}{{\phi}^{\gamma}}.
\end{equation}

Numerically we determine ${{C}_{0}}=5.9$ and $\gamma=1.82$. For large $\Delta$
the exponential in Eq.~(\ref{TC}) is too abrupt and there are no real
solutions. For lower $\Delta$ there is a domain which admits real solution,
which is bigger the lower is $\Delta$, so that there can be solutions of
finite magnetization at low temperature. For low $\phi$, the only solution for
$M$ is 0 if $J$ approaches 1.

With respect to the order parameter related with neutrality, see
Figure~\ref{WSframe} (right), when $\Delta= 0$ and $M = 0$ all solutions of
$1-x$ are null at any temperature, i.e., all the agents are neutral. Comparing
this figure with Figure~\ref{RGframe} we can see that while neutrality in the
fully disordered solution is approximately the same in both models, in
Watts-Strogatz the neutrality is narrower, i.e., the same neutrality
corresponds to lower magnetization.

\begin{figure}[ptb]
\centering
\includegraphics[scale=0.6]{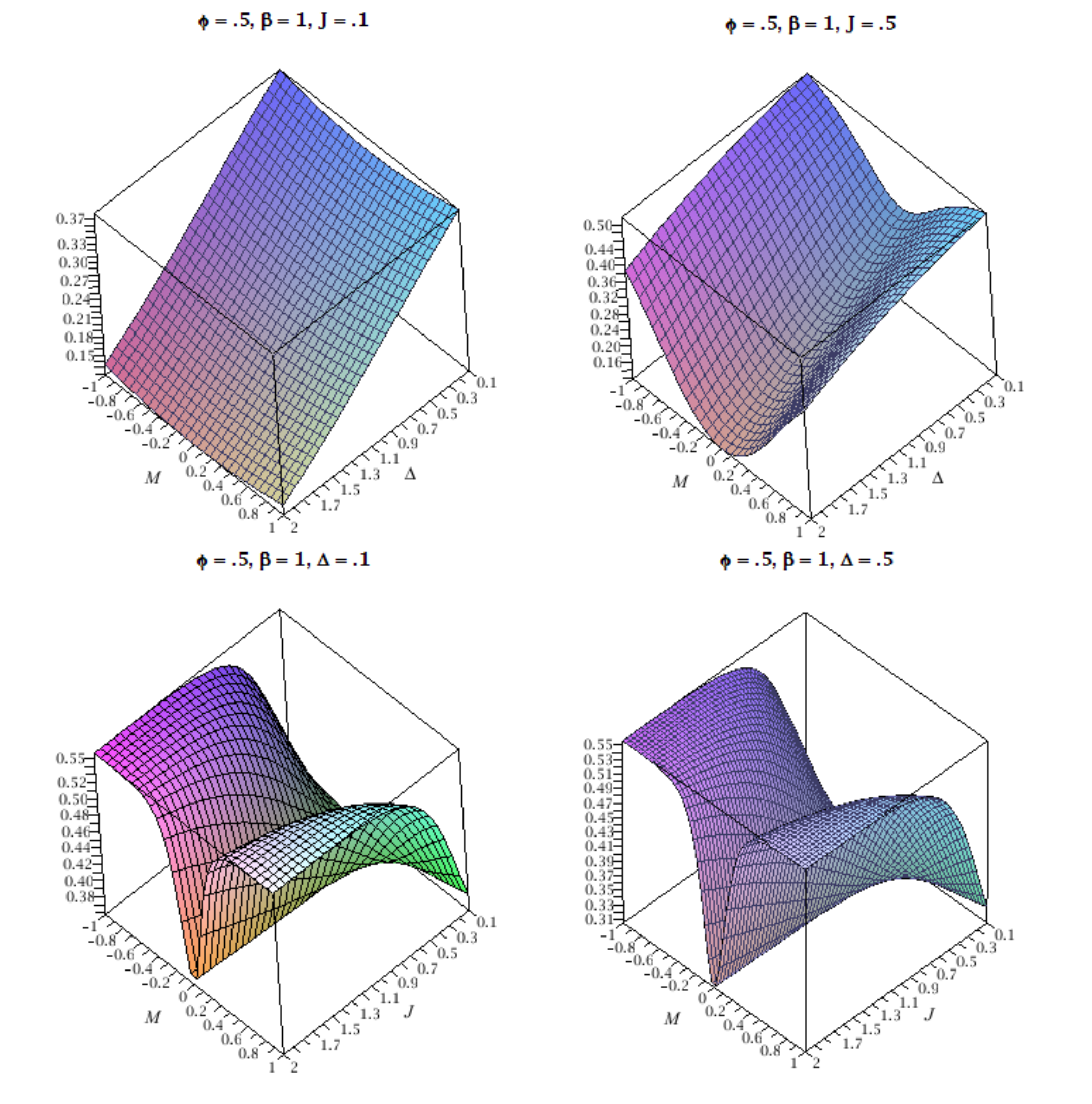}\caption{(Color online). Effect
of varying $J$ and $\Delta$ on the neutrality, represented in $z$ axis, in the
Watts-Strogatz substrate.}%
\label{ws3d}%
\end{figure}

The effect of varying it and $J$ can be seen in Figure~\ref{ws3d}, which
represents neutrality for a fixed $\phi$. For a fixed $\Delta$, increasing $J$
reduces the region of high neutrality. Comparing to the random graph, the
addition of shortcuts increases the importance of the strength of
interactions. If the message can be conveyed to other distant groups, its
repercussion is bigger and overcomes the propensity to immobility or
neutrality. This influence is stronger than increasing $\Delta$ while keeping
$J$ constant.

\subsection{Newman Substrate}

\label{Newman} Let's analyze now the variant proposed by Newman to the
Watts-Strogatz model, as described in Section~(\ref{WS}), and remember that to
build this model, Newman starts from a regular unidimensional grid in which
each node is connected to $2z$ nearest neighbors, then adds a link with
probability $\phi$ to a randomly chosen pair of nodes.

The calculations are simpler in this case. We start initially from a degree
$2z$ without shortcuts, so that we have $nz$ links. If now we add a shortcut
with probability $\phi$ we will have $nz\phi$ shortcuts, $2nz\phi$ shortcut
ends from which $2z\phi$ link by average to a given node. The number $s$ of
shortcuts which end in a given node follows a Poisson distribution with mean
$2z\phi$:
\begin{equation}
P(s)={{e}^{-2z\phi}}\frac{{{(2z\phi)}^{s}}}{s!}.
\end{equation}
As the degree of a node is $q=2z+s$ the distribution is
\begin{equation}
\label{PqN}P(q,z,\phi)={{e}^{-2z\phi}}\frac{{{(2z\phi)}^{q-2z}}}{(q-2z)!}%
\end{equation}
for $q\ge2z$, and $P(q,z,\phi)=0$ for $q<2z$.

\begin{figure}[ptb]
\centering
\includegraphics[scale=0.5]{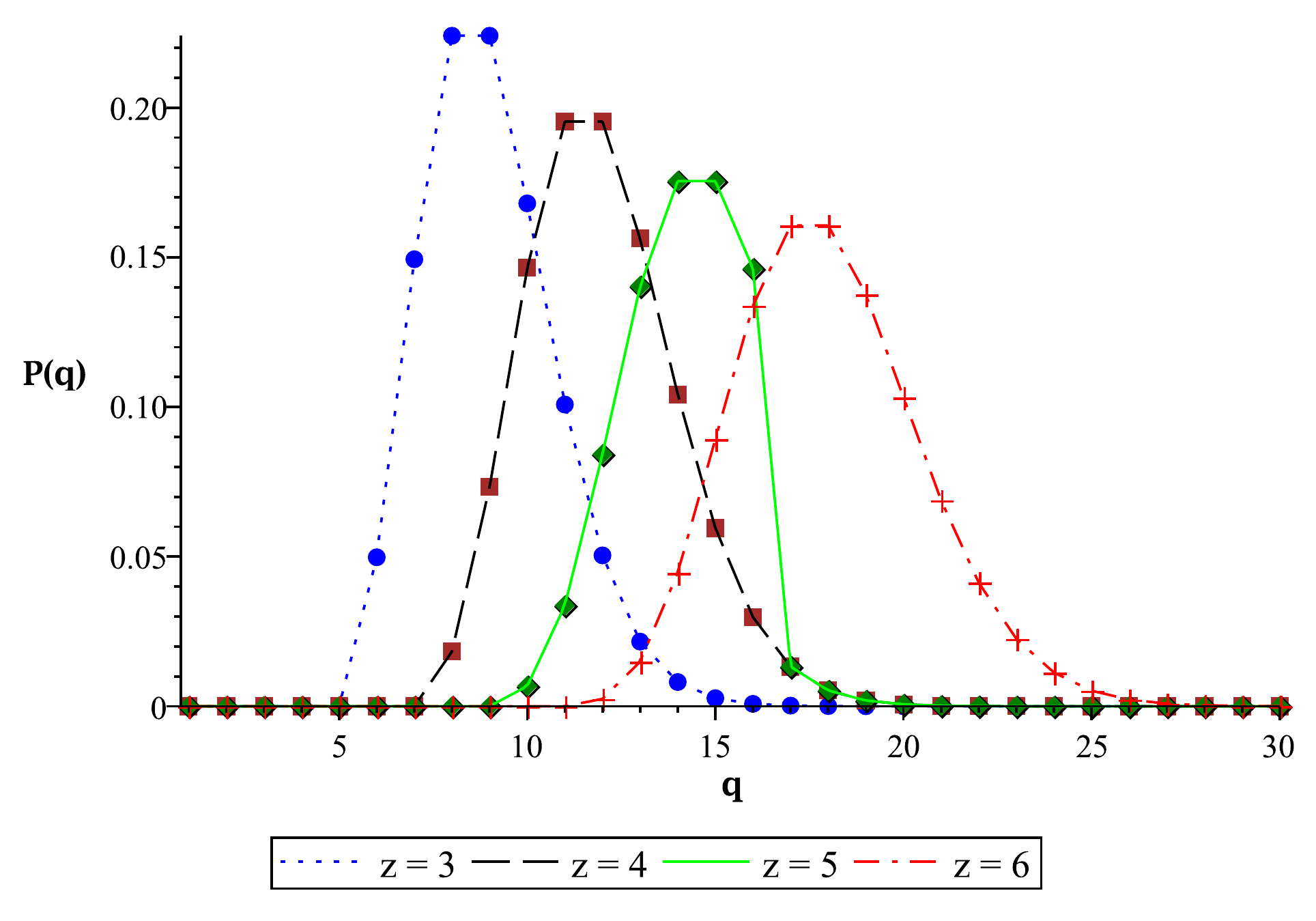}\caption{(Color online).
Connectivity in Newman substrate for different levels of neighborhood.}%
\label{NKz}%
\end{figure}

Figure~\ref{NKz} shows the connectivity distribution of this model for
different levels of next neighbors.

Observing the rapid drop at large degrees in the distribution for $z=3$ it is
apparent that we can safely use a cutoff of 20. Then the Annealed Network
approach having substituted the degree distribution Eq.~(\ref{PqN}) in
Eq.~(\ref{M}) yields for the order parameters
\begin{equation}
\label{M2}M=\sum\limits_{q=6}^{20}{\left\{  \frac{q}{\left\langle q
\right\rangle }\cdot P(q,z,\phi)\cdot\frac{2\sinh\beta JqM}{\exp\left(
\beta\Delta\right)  +2\cosh\beta JqM} \right\}  }%
\end{equation}
and
\begin{equation}
1-x=\sum\limits_{q=6}^{20}{\left\{  \frac{q}{\left\langle q \right\rangle
}\cdot P(q,z,\phi)\cdot\frac{2\cosh\beta JqM}{\exp\left(  \beta\Delta\right)
+2\cosh\beta JqM} \right\}  }.
\end{equation}

As in former models $M=0$ is always a solution, and there is a limiting value
of $\Delta$ for which it is unique, so that below this value only exists the
disordered state and above there are more finite solutions.

\begin{figure}[ptb]
\centering
\includegraphics[scale=0.3]{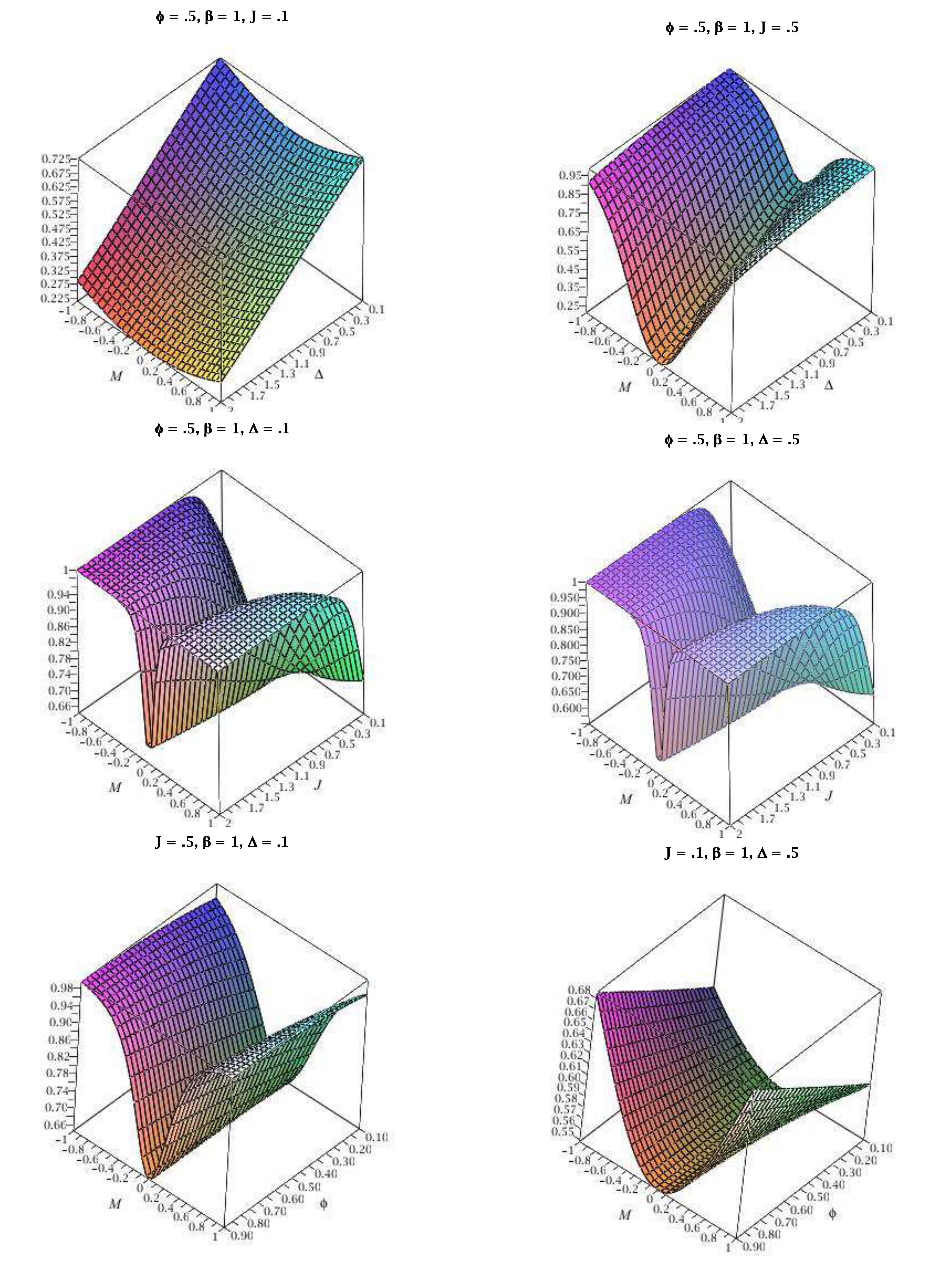}\caption{(Color online).
Effect of varying $J$, $\Delta$ and $\phi$ on the neutrality, represented in
$z$ axis, in the Newman substrate.}%
\label{newman3d}%
\end{figure}

For large $\Delta$ the exponential in Eq.~(\ref{TC2}) is too abrupt and there
is not a real value for critical temperature, implying that finite
magnetization at low temperature is not possible. On the contrary, for low
$\Delta$ there is a domain which admits real solution, bigger the lower is
$\Delta$, and in this case finite magnetization at low temperature is
possible. The effect of varying $\Delta$ and $J$ can be seen in
Figure~\ref{newman3d}. Comparing this dependence with the corresponding one
shown in Figure~\ref{ws3d}, one can see that both cases have an inverse
dependence on the average degree, being larger in the Newman%
%TCIMACRO{\U{b4}}%
%BeginExpansion
\'{}%
%EndExpansion
s case. One can also observe that the increase of $\Delta$ leads to higher
values for the neutrality, in correspondence with the results in
\cite{Schelling2} where higher values of the same parameter (interpreted as
urban attractiveness) lead to an unwelcoming environment as more and more
agents leave.

For low $\phi$, the only solution for $M$ is 0 if $J$ approaches 1. The
critical temperature is found graphically as before, looking at the
intersections of $y=M$ and the rhs of Eq.~(\ref{M2}) to get the critical value
of $\beta$ at which a non-zero value of $M$ is possible:
\begin{equation}
\frac{C{{\beta}_{c}}J}{{{e}^{{{\beta}_{c}}\Delta}}+2}=1.\label{TC2}%
\end{equation}
where $C$ is a constant with an approximate linear dependence on the
reconnection probability $\phi$ of the form
\begin{equation}
C={{C}_{0}}+\alpha\phi.
\end{equation}
and numerically we obtain ${{C}_{0}}=12.1,\alpha=12.9$.

In order to understand the effect of the topology on the model behavior, we
have performed an additional analysis for different values of the probability
of connection $\phi$, shown in Figure~\ref{newman3d}. As can be seen, although
the quantitative behavior vary upon the probability distribution of the
shortcut connections, the qualitative behavior does not depend very much on
the detail. This observation also holds when comparing the results with the
Mean Field (MF) results of the Blume-Capel model on a regular lattice
($\phi\rightarrow1$), although one observes a slight enhancement of the area
of the ordered phase when decreasing the value of $\phi$.

\subsubsection{Phase Diagram}

In this section we will study the Newman substrate in further detail. The
reason is that we think that while it seems realistic to establish new links
between agents, it is not so to destroy existing ones. This could be the case
of the break of diplomatic relations in case of war. But then, the war is not
subject of study in this paper.

\begin{figure}[tbh]
\centering
\includegraphics[scale=0.4]{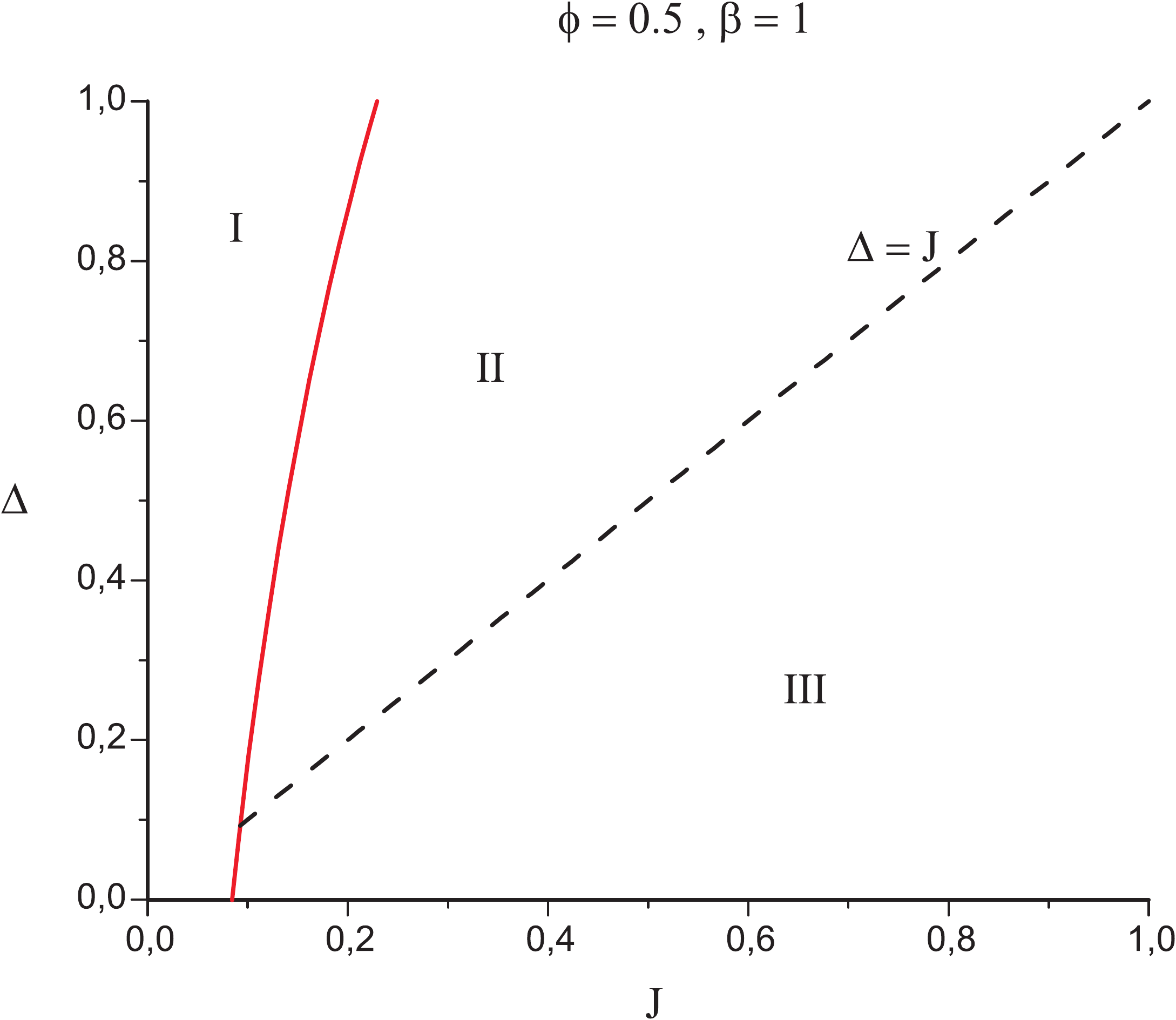}\caption{(Color online).
Phase diagram in Newman substrate for $z=3$.}%
\label{DF}%
\end{figure}

Figure~\ref{DF} shows the phase diagram for $z=3$. We can see in it three main
regions. The solid line separates a disordered phase with $M=0$ on the left
from an ordered phase with $M\neq0$ on the right. The line $\Delta=J$
separates the region of finite magnetization with an appreciable concentration
of neutral agents from another in which alignment is so strong that there are
almost no neutrals in temperatures for which fluctuations are not important.

If $J$ is high enough, one of the options prevails over the other and the
existence of neutrality is strongly influenced by the value of $\Delta$, being
higher the higher the temperature. For $\Delta$ below that line, the neutral
component is negligible.

For this case, a comparison with the results obtained by using the Ising model
is interesting \cite{Nadal2,Bouchaud}. For values of the parameter $\Delta$
below the dashed line shown in Fig. 8, the social influence is stronger, and
the agents make a decision in accordance with the classical Ising model in
absence of external field (absence of personal inclination in
\cite{Nadal2,Bouchaud}).

In our model, the effect of network topology is manifested on its effect on
the connectivity of the reconnection probability. A high value of $\phi$
favors alignment, as it increases connectivity and hence the amount of
interactions for the same $J$ so that the null magnetization phase occurs for
smaller J.

\paragraph{\textbf{Region I}}

\textbf{$J$ small, $\Delta$ big. Null magnetization with finite neutrality.}
In this region the competing options do not appeal the agents enough so as to
produce a clear prevalence. Due to the low intensity of interaction in
comparison with the propensity to remain neutral, very few agents decide to
align, in such a way that magnetization is kept low because the alignment is symmetric.

In politics this is like apathy or general disappointment with the main
contending options produced by these not being able to communicate, or failing
to convey the right message. In short, abstention wins.

\begin{figure}[ptb]
\centering
\includegraphics[width=\textwidth]{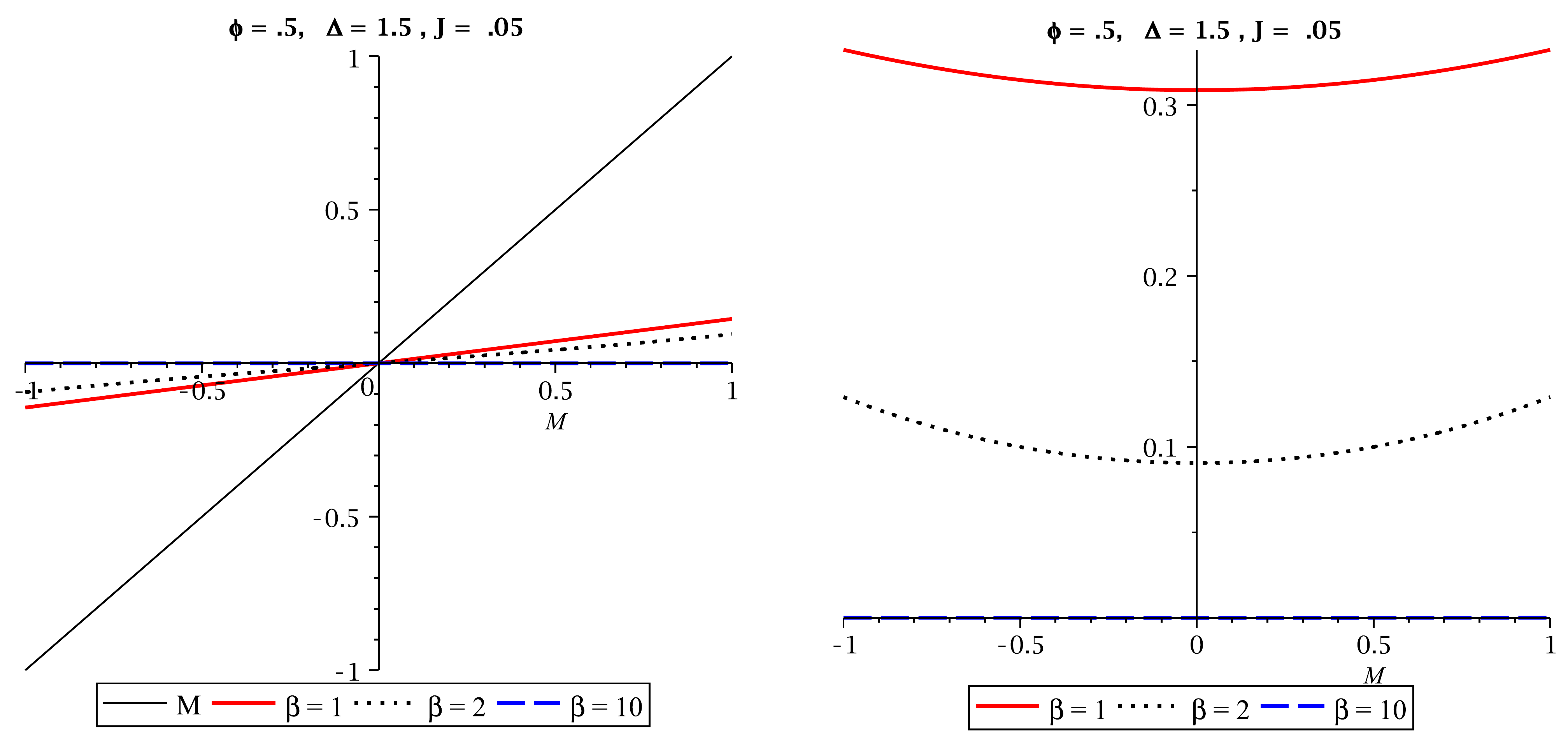}\caption{(Color
online). Solutions for magnetization (left) and neutrality (right) in Region
I.}%
\label{RIframe}%
\end{figure}

Figure~(\ref{RIframe}) displays graphs of solutions for magnetization and
neutrality for different values of the control parameters in this region.

\paragraph{\textbf{Region II}}

\textbf{$J$ big, $\Delta>J$. Finite magnetization with finite neutrality.} In
this region, the strength of interactions is such that the advantages of one
of the contending options are clearly perceived, but however the effect of
$\Delta$ favouring neutrality is stronger.

If $\Delta\gg J$ neutrality prevails over other options in equilibrium states
although the magnetization is not null. People behave in a conservative way,
not aligning, not making investments, not expending, but more meaningfully,
not deciding. Politic parties don't excite the society and the options voted
win with notorious abstention.

\begin{figure}[ptb]
\centering
\includegraphics[width=\textwidth]{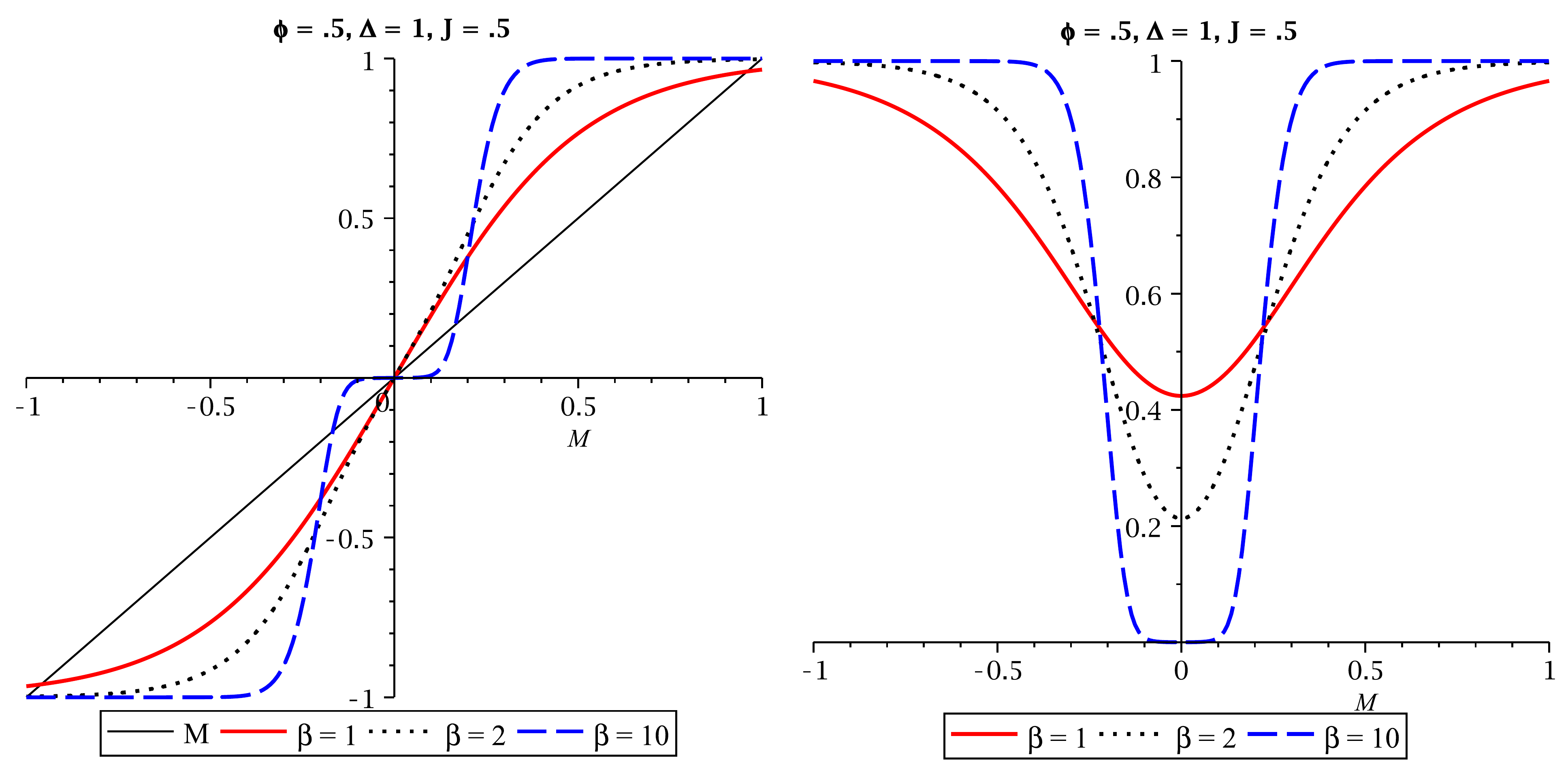}\caption{(Color
online). Solutions for magnetization (left) and neutrality (right) in Region
II.}%
\label{RIIframe}%
\end{figure}

Figure~(\ref{RIIframe}) displays graphs of solutions for magnetization and
neutrality for different values of the control parameters in this region.

\paragraph{\textbf{Region III}}

\textbf{$J$ big, $\Delta<J$. Finite magnetization with low neutrality.} This
region corresponds to the clear predominance of one of the options, whose
advantages the agents perceive clearly, so that they feel motivated for a
sharp alignment. The theoretical solution corresponds to high magnetization
with null neutrality, with a prevailing option which depends on the initial state.

In politics, this is the situation of absolute majority of a party.
Commercially, one of the options is the preferred choice by majority.

\begin{figure}[ptb]
\centering
\includegraphics[width=\textwidth]{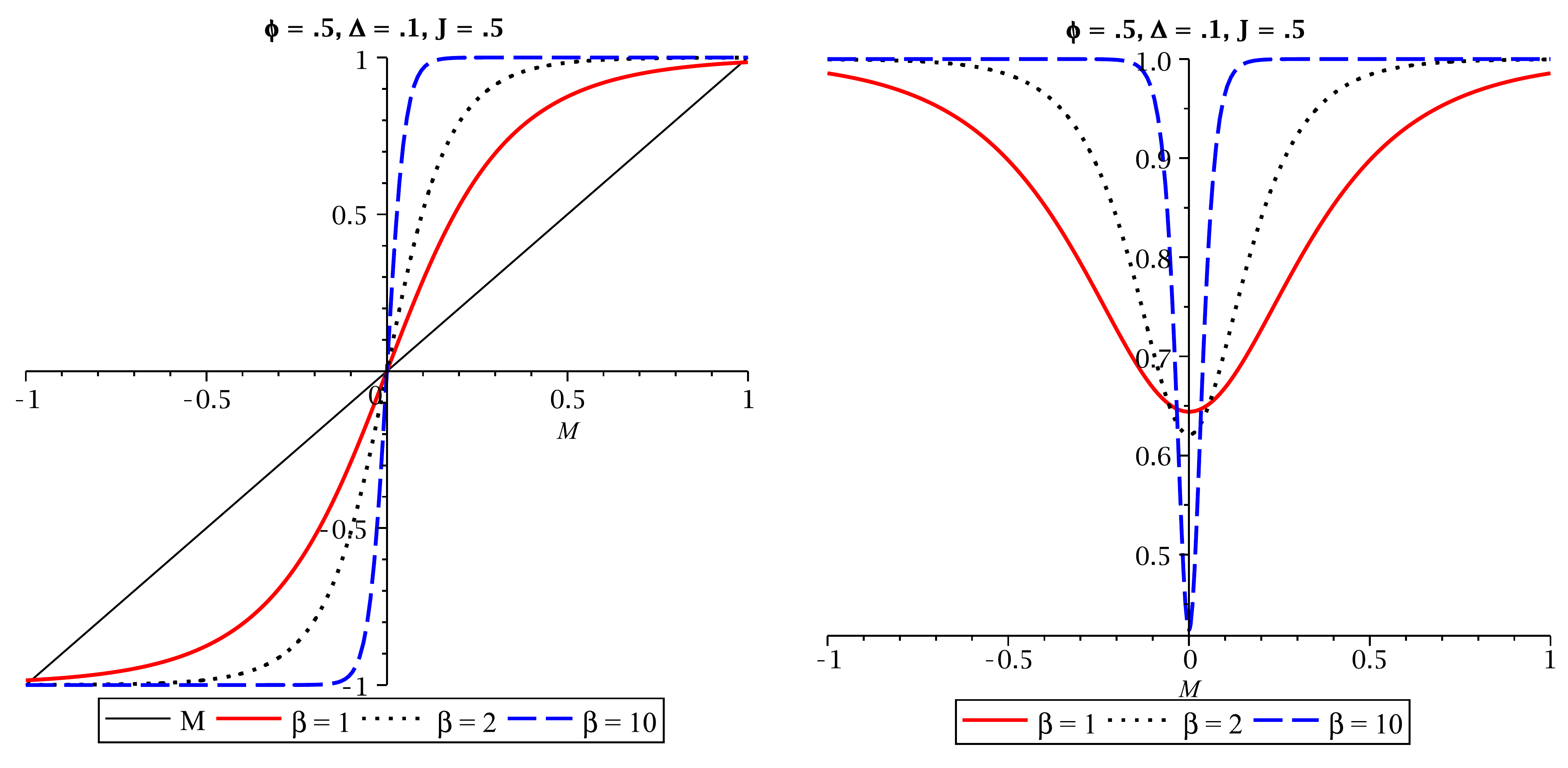}\caption{(Color
online). Solutions for magnetization (left) and neutrality (right) in Region
III.}%
\label{RIIIframe}%
\end{figure}

Figure~(\ref{RIIIframe}) displays graphs of solutions for magnetization and
neutrality for different values of the control parameters in this region.

\section{Discussion}

In this paper we have studied a 3-states magnetic model residing on three
topologically different substrates, establishing sociophysical analogies for
the control parameters.

Local interaction is determined by the coupling constant $J$. Socially
corresponds to an exchange of information, which happens formally as
advertisement, or by direct contact. In politics is related to commercial or
diplomatic treatises. The anisotropy term $\Delta$, which in the Blume-Capel
model is associated to the chemical potentials of transformation from one
species to another, represents in our model the energetic cost to pass from
neutrality to alignment or vice versa.

Global interaction, in turn, is determined by the reconnection probability
$\phi$. It defines the number of linked agents and hence the connectivity.
Therefore, it has only any impact in the Newman substrate, because
connectivity is fixed in the others. The effect of this parameter is
manifested as a decrease in neutrality, almost unseen in the Watts-Strogatz
substrate. This seems to indicate that isolation favors neutrality and is
appreciable in the extremes, i.e., for the finite solutions of magnetization.
In sociophysics, this means that the polarization is more intense for higher
connectivity, which can be interpreted perhaps paradoxically as if a better
communication favors discrepancy.

Neutrality is strongly conditioned by the anisotropy constant $\Delta$. Its
effect is apparent in the critical temperature and is the key control
parameter in our model, as the magnetization is only relevant to distinguish
between ordered and disordered states. For a small value of $\Delta$ the
system tends to distribute in the two contending options, but if $\Delta$ is
too high, there is no chance for consensus in any of the options, so that all
the agents are indifferent (neutral). Socially, the higher interaction
strength due to the bigger connectivity implies better communication and
information, thus canceling the effects of indifference towards the advantages
of any of the contending options. $\Delta$ can be regarded as opposition to
the main options in a two-party politic system, or in other words, abstention
with sympathy for minority options. In diplomacy it represents the advantages
of not aligning with any of two conflicting options. When it comes to study
customer options, it is the savings of not buying anything.

The results presented in this work can be compared with those obtained by
using the RFIM \cite{Nadal2, Bouchaud}. For low values of the parameter
$\Delta$ the system tends to align in a binary-like scenario, thus leading to
well known MF results. It is also interesting to point out that there is a
good qualitative agreement with the MF results of the Blume-Capel model on a
regular lattice, although the quantitative comparison depends on the
probability of connections.

The analogy between the BEG model and his variant of Blume-Capel spin-1 model
with the Schelling's model \cite{Schelling1} and its generalization for an
open city \cite{Schelling2} permits to relate variables such as neutrality (in
opinion formation) with empty location (in Schelling's models), showing
similar behaviors in the sense that higher values of the anisotropy parameter
$\Delta$ --the urban attractiveness in \cite{Schelling2}-- leads to higher neutrality.

With respect to the substrate, we found that in Newman substrate the value of
neutrality is less than in Watts-Strogatz for the same value of $\Delta$ and
$J$. This appears to be due to the higher connectivity, because in the Newman
model links are just added without deleting previously existing ones.

For high reconnection probability, Watts-Strogatz substrate approaches the
random graph, as its degree distribution tends to Poisson for reconnection
probability near to 1. Logically, for low reconnection probability, Newman and
random graph models are more similar as both approach the regular grid
behavior (compare Figures ~\ref{RGframe} and ~\ref{WSframe} with the different
regions in the phase space of Newman substrate model). We think that the later
represents better more real networks in which an agent has closer relations
with its neighbors. Vicinity must be understood here as imposed by the system
considered: work mates, neighboring countries, etc., extending links equally
intimate with geographically far agents, so being neighbors in this concept of
distance. We also feel this model is more realistic as, being normal to
establish new relations, it is not so normal that these replace existing
relations. Moreover, the connectivity in this model varies with the
reconnection probability, which allows to study more different situations.

In the Newman substrate the critical temperature is lower than in
Watts-Strogatz, while below this temperature finite solutions for
magnetization are higher. The meaning of the existence of a critical
temperature is that a very high agitation, either thermal as in the magnetic
model, or social, prevents the agents to settle in a fixed state. In this
paper we interpret the temperature as a finite probability of doing a change
of state which is not energetically prescribed. In sociophysical terms, an
agent will take a decision which is not the most interesting.

To summarize, if we model society as made of individuals which can adopt one
of two possible options or none of these, we conclude that the higher
connectivity will increase the level of alignment at the expenses of
neutrality. The RFIM provides a unifying framework to account for many
collective socioeconomic phenomena that lead to endogenous ruptures and
crisis. Moreover, it has been quite successfully used to analyze
quantitatively real data \cite{Michard-Bouchaud}. However, we think that the
three-states model considered in this work is a good approach for modeling
more complex decision processes. A straightforward comparison between our
model and real situations is difficult, but the World War II , the Cold War
\cite{Galam2} or the fragmentation of the ex-Yugoslavia \cite{Florian} might
be good real examples of application.

\smallskip In order to test our analytical results and to apply our model to
more complicated and realistic situations, we have developed an agent-based
computer simulation tool. The simulator is based on the Netlogo v5.0.2 and
implements the Uri Wilenski algorithm for the generation of different
substrates ~\cite{Watts}. To identify small worlds, the average path length
and clustering coefficient of the network are calculated and plotted. All
control parameters can be set up, and both numerical and graphical outputs are
available. With it, we have confirmed that the outcomes correspond to stable
configurations in a broad range of the control parameters, in good agreement
with the analytical results. Moreover, as a first step towards more realistic
cases, we have applied our model to the study of the coalitions during the
Cold War using real data obtained from The Correlates of War Project (COW)
\cite{COW}. For that purpose, we chose 1955 as the oldest year, for which good
commerce real data are known, and used the correspondent commerce flows
between different countries to fix the coupling interactions in the context of
our model. We ran simulations in Region III of the phase diagram (see
Figure~\ref{DF}). Our preliminary results show the grouping of two big
coalitions, with a good resemblance to the NATO and Warsaw Pact, as well as
the group of neutral countries. More detailed work in this direction is in progress.

\section*{Acknowledgements}

The authors thank H. Chamati and N. Tonchev for fruitful discussions, M.
Jimenez for the preprocessing of real data, and the anonymous referees for
their interesting comments and suggestions.

\smallskip

%\begin{appendices}

\section*{Appendix. Order parameters in the Annealed Network Approximation}

If $P({{S}_{i}})$ is the probability of a site having the spin ${{S}_{i}}$ we
have
\begin{align}
{{m}_{i}}  &  =\left\langle {{S}_{i}}\right\rangle =\frac{\sum
\limits_{\left\{  S\right\}  }{{{S}_{i}}P({{S}_{i}})}}{\sum\limits_{\left\{
S\right\}  }{\exp(-\beta{{E}_{i}})}}=\frac{{{e}^{-\beta{{E}_{+}}}}%
-{{e}^{-\beta{{E}_{-}}}}}{{{e}^{-\beta{{E}_{+}}}}+{{e}^{-\beta{{E}_{0}}}}%
+{{e}^{-\beta E-}}}=\nonumber\\
&  =\frac{\exp\left\{  \beta\left[  \sum\limits_{i\neq j}{\left(  {{J}_{ij}%
}{{m}_{j}}\right)  -\Delta}\right]  \right\}  -\exp\left\{  -\beta\left[
\sum\limits_{i\neq j}{\left(  {{J}_{ij}}{{m}_{j}}\right)  +\Delta}\right]
\right\}  }{\exp\left\{  \beta\left[  \sum\limits_{i\neq j}{\left(  {{J}_{ij}%
}{{m}_{j}}\right)  -\Delta}\right]  \right\}  +\exp\left\{  -\beta\left[
\sum\limits_{i\neq j}{\left(  {{J}_{ij}}{{m}_{j}}\right)  +\Delta}\right]
\right\}  +1}\nonumber\\
&  =\frac{\exp\left\{  \beta\sum\limits_{i\neq j}{\left(  {{J}_{ij}}{{m}_{j}%
}\right)  }\right\}  -\exp\left\{  -\beta\sum\limits_{i\neq j}{\left(
{{J}_{ij}}{{m}_{j}}\right)  }\right\}  }{\exp\left\{  \beta\sum\limits_{i\neq
j}{\left(  {{J}_{ij}}{{m}_{j}}\right)  }\right\}  +\exp\left\{  -\beta
\sum\limits_{i\neq j}{\left(  {{J}_{ij}}{{m}_{j}}\right)  }\right\}
+\exp(\beta\Delta)} \tag{A.1}%
\end{align}

and finally
\begin{equation}
{{m}_{i}}=\frac{2\sinh\left[  \beta\sum\limits_{i\neq j}{\left(  {{J}_{ij}%
}{{m}_{j}}\right)  }\right]  }{{{e}^{\beta\Delta}}+2\cosh\left[  \beta
\sum\limits_{i\neq j}{\left(  {{J}_{ij}}{{m}_{j}}\right)  }\right]  }.
\tag{A.2}%
\end{equation}

Note the formal similarity with Eq.(\ref{BCM}). Proceeding in the same way for
the neutrality we have%
\begin{align}
1-x  &  =\left\langle S_{i}^{2}\right\rangle =\frac{\sum\limits_{\left\{
S\right\}  }{S_{i}^{2}P({{S}_{i}})}}{\sum\limits_{\left\{  S\right\}  }%
{\exp(-\beta{{E}_{i}})}}=\frac{{{e}^{-\beta{{E}_{+}}}}+{{e}^{-\beta{{E}_{-}}}%
}}{{{e}^{-\beta{{E}_{+}}}}+{{e}^{-\beta{{E}_{0}}}}+{{e}^{-\beta E-}}%
}=\nonumber\\
&  =\frac{\exp\left\{  \beta\left[  \sum\limits_{i\neq j}{\left(  {{J}_{ij}%
}{{m}_{j}}\right)  -\Delta}\right]  \right\}  +\exp\left\{  -\beta\left[
\sum\limits_{i\neq j}{\left(  {{J}_{ij}}{{m}_{j}}\right)  +\Delta}\right]
\right\}  }{\exp\left\{  \beta\left[  \sum\limits_{i\neq j}{\left(  {{J}_{ij}%
}{{m}_{j}}\right)  -\Delta}\right]  \right\}  +\exp\left\{  -\beta\left[
\sum\limits_{i\neq j}{\left(  {{J}_{ij}}{{m}_{j}}\right)  +\Delta}\right]
\right\}  +1}=\nonumber\\
&  =\frac{2\cosh\left[  \beta\sum\limits_{i\neq j}{\left(  {{J}_{ij}}{{m}_{j}%
}\right)  }\right]  }{{{e}^{\beta\Delta}}+2\cosh\left[  \beta\sum
\limits_{i\neq j}{\left(  {{J}_{ij}}{{m}_{j}}\right)  }\right]  } \tag{A.3}%
\end{align}

Using now the Annealed Network Approximation described in Section 3 we get the
following approximations for the order parameters%

\begin{equation}
{{m}_{i}}=\frac{2\sinh\left[  \frac{\beta J{{q}_{i}}}{\left\langle
q\right\rangle N}\sum\limits_{j}{{{q}_{j}}{{m}_{j}}}\right]  }{{{e}%
^{\beta\Delta}}+2\cosh\left[  \frac{\beta J{{q}_{i}}}{\left\langle
q\right\rangle N}\sum\limits_{j}{{{q}_{j}}{{m}_{j}}}\right]  }=\frac
{2\sinh\beta J{{q}_{i}}M}{{{e}^{\beta\Delta}}+2\cosh\left[  \beta J{{q}_{i}%
}M\right]  } \tag{A.4}%
\end{equation}
and%
\begin{align}
1-x  &  =\frac{2\cosh\left[  \beta\sum\limits_{i\neq j}{\left(  {{J}_{ij}}%
{{m}_{j}}\right)  }\right]  }{{{e}^{\beta\Delta}}+2\cosh\left[  \beta
\sum\limits_{i\neq j}{\left(  {{J}_{ij}}{{m}_{j}}\right)  }\right]  }%
=\frac{2\cosh\left[  \frac{\beta J{{q}_{i}}}{\left\langle q\right\rangle
N}\sum\limits_{j}{{{q}_{j}}{{m}_{j}}}\right]  }{{{e}^{\beta\Delta}}%
+2\cosh\left[  \frac{\beta J{{q}_{i}}}{\left\langle q\right\rangle N}%
\sum\limits_{j}{{{q}_{j}}{{m}_{j}}}\right]  }=\nonumber\\
&  =\frac{2\cosh\beta J{{q}_{i}}M}{{{e}^{\beta\Delta}}+2\cosh\left[  \beta
J{{q}_{i}}M\right]  } \tag{A.5}%
\end{align}

%\end{appendices}

\end{document}